\documentclass[10pt,a4paper]{article}
\usepackage{lrec2026} % use [review] for anonymized submission

\usepackage{multirow}
\usepackage{placeins}
\usepackage{pbox}
\usepackage{colortbl}
\usepackage{graphicx}
\usepackage{xspace}
\usepackage{tabularx}
\usepackage{tabulary}
\usepackage{amsmath}
\usepackage{booktabs}
\usepackage{listings}
\usepackage[most]{tcolorbox}
\usepackage{cuted}
\usepackage{xcolor}
\usepackage{kotex}
\usepackage{pifont}
\usepackage{dirtytalk}
\usepackage{algorithm}
\usepackage{algpseudocode}
\usepackage[perpage]{footmisc}
\usepackage{hyperref}
\usepackage{makecell}
\usepackage{array}

\newtcolorbox{instructionsbox}[1][]{
  colframe=cyan!75!black,
  colback=green!5!white,
  coltitle=black,
  title=#1,
  rounded corners,
  boxrule=0.5mm,
  boxsep=5pt,
  toptitle=1mm,
  bottomtitle=1mm,
  left=0pt,
  right=0pt,
  top=0pt,
  bottom=0pt,
  fonttitle=\bfseries
}

\newcommand{\hret}{\textsc{HRET}\xspace}

\title{Redefining Evaluation Standards: A Unified Framework for Evaluating the Korean Capabilities of Language Models}

\name{\parbox{\textwidth}{\centering\bfseries\large
  Hanwool Lee\textsuperscript{1,2,*},
  Dasol Choi\textsuperscript{1,*},
  Sooyong Kim\textsuperscript{3},
  Ilgyun Jeong\textsuperscript{4},
  Sangwon Baek\textsuperscript{5},
  Guijin Son\textsuperscript{2},
  Inseon Hwang\textsuperscript{6},
  Naeun Lee\textsuperscript{7},
  Seunghyeok Hong\textsuperscript{8,†}
\vspace{3pt}}}
\address{%
\textsuperscript{1}AIM Intelligence\quad
\textsuperscript{2}Seoul National University\quad
\textsuperscript{3}A.I.MATICS\quad
\textsuperscript{4}TigerCompany\quad
\textsuperscript{5}Catius \\
\textsuperscript{6}National Assembly of Korea\quad
\textsuperscript{7}Coupang\quad
\textsuperscript{8}Hankuk University of Foreign Studies \\
\texttt{gksdnf424@gmail.com}
}

\abstract{
Recent advances in Korean large language models (LLMs) have produced numerous benchmarks and evaluation methods, yet inconsistent protocols lead to performance gaps of up to 10 percentage points across institutions. These discrepancies arise from fragmented practices in prompt templates, inference settings, and evaluation criteria, making reliable comparison difficult.
We introduce \textbf{HRET (Haerae Evaluation Toolkit)}, an open-source, registry-based framework that unifies Korean LLM evaluation. HRET standardizes data ingestion, inference, and reporting pipelines to ensure reproducibility, while remaining flexible through modular integration of major Korean benchmarks, multiple inference backends, and diverse evaluation methods (string matching, log-likelihood, mathematical verification, and LLM-as-judge). 
Beyond standard accuracy metrics, HRET provides Korean-specific diagnostic analyses—morphology-aware Type–Token Ratio for lexical diversity and systematic keyword-omission detection for conceptual coverage—offering linguistic insights beyond conventional evaluation. 
Across diverse hardware and backend configurations, HRET yields score variations below 0.003 (vs.\ 0.1+ pp gaps reported previously), demonstrating its effectiveness in eliminating evaluation discrepancies and establishing a reproducible standard for Korean LLM benchmarking. HRET is publicly available at \url{https://github.com/HAE-RAE/haerae-evaluation-toolkit}.\\ \newline
\Keywords{Korean NLP, Evaluation Framework, Reproducibility, Benchmarking, LLM-as-Judge}
}

\begin{document}

\maketitleabstract

\renewcommand{\thefootnote}{\fnsymbol{footnote}}
\footnotetext[1]{Equal contribution.}
\footnotetext[2]{Corresponding author.}
\renewcommand{\thefootnote}{\arabic{footnote}}

% ---------------- Main content (content preserved) ----------------

\begin{figure*}[h]
    \centering
    \includegraphics[width=0.99\textwidth]{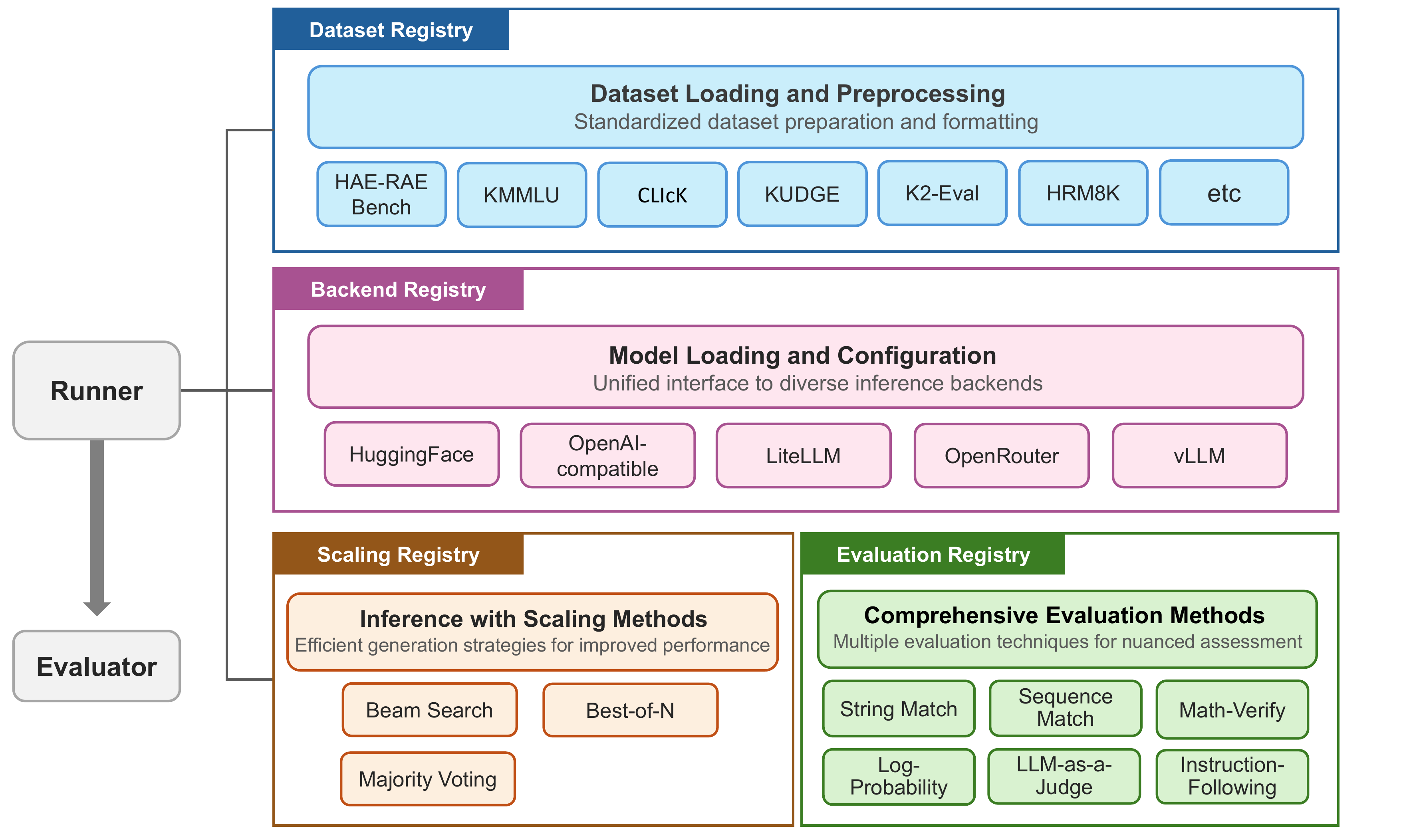}
    \caption{Architecture overview of HRET, showing the registry-based modular design that integrates datasets, models, inference strategies, and evaluation methods through a unified pipeline.}
    \label{fig:arch}
\end{figure*}

\section{Introduction}
Recent advances in large language models (LLMs) demand rigorous and reproducible evaluation, yet real-world benchmarking often diverges from official protocols~\cite{liang2022holistic, eval-harness}. In Korean NLP, identical models have shown discrepancies of up to 10 percentage points across institutions~\cite{exaone, kanana}, largely due to variations in prompt templates, inference settings, and evaluation criteria. Such inconsistency undermines fair comparison and slows progress in the field.

The reproducibility crisis in Korean LLM evaluation arises from two main factors. First, practitioners face a fragmented landscape: even widely used benchmarks such as KMMLU~\cite{kmmlu} in lm-eval-harness are rarely applied without modification. Researchers frequently augment them with few-shot examples~\cite{brown2020language}, chain-of-thought prompts, or self-consistency voting~\cite{wang2023selfconsistencyimproveschainthought} to better capture reasoning. Second, Korean's linguistic characteristics—rich morphology, flexible word order, and frequent case-marker omission—introduce tokenization and normalization differences that materially affect scores. Without standardized handling of these phenomena, reproducibility remains elusive.

While general-purpose frameworks such as lm-evaluation-harness~\cite{eval-harness} and DeepEval~\cite{deepeval} have advanced English evaluation, they lack native support for Korean linguistic phenomena, requiring significant manual adaptation that often reintroduces the inconsistencies they aim to eliminate.

We introduce \textbf{HRET} (\textbf{Haerae Evaluation Toolkit}), an open-source, registry-based framework that unifies Korean LLM assessment. HRET standardizes evaluation workflows—data ingestion, inference, and result reporting—while remaining flexible through modular integration of major Korean benchmarks, multiple inference backends, and diverse evaluation methods (string matching, log-likelihood, mathematical verification, and LLM-as-judge). It further enforces language consistency to penalize non-Korean outputs and incorporates Korean-specific diagnostics: morphology-aware Type–Token Ratio for lexical diversity and systematic keyword-omission detection for conceptual coverage. Beyond its pipeline-level design, HRET provides a high-level decorator-style API inspired by DeepEval, enabling one-line evaluation, multi-model benchmarking, and seamless MLOps tracking. We validate HRET across diverse hardware and backends, showing performance variations below 0.003 standard deviation—an order of magnitude smaller than prior reports—demonstrating a practical path toward reproducible, adaptable Korean LLM evaluation.

\begin{table*}[h]
\centering
\small
\renewcommand{\arraystretch}{1.0}
\resizebox{0.85\textwidth}{!}{
\begin{tabular}{llcccccc}
\toprule
\textbf{Report} & \textbf{Model} & \multicolumn{5}{c}{\textbf{Datasets}} \\
\cmidrule(lr){3-7}
 & & \textbf{GSM8K} & \textbf{MMLU} & \textbf{KMMLU} & \textbf{HumanEval} & \textbf{MBPP} \\
\midrule
\multirow{2}{*}{\shortstack[l]{EXAONE 3.5 Report\\\cite{exaone}}}
 & Qwen2.5-32B  & 92.00  & 81.40  & 62.10  & 89.00  & 88.90  \\
 & Gemma2-27B   & 84.20  & 74.80  & 53.80  & 79.30  & 80.70  \\
\midrule
\multirow{2}{*}{\shortstack[l]{Kanana Report\\\cite{kanana}}}
 & Qwen2.5-32B  & 95.30  & 84.40  & 59.37  & 82.32  & 71.96  \\
 & Gemma2-27B   & 91.05  & 78.30  & 49.98  & 70.12  & 70.90  \\
\bottomrule
\end{tabular}
}
\caption{Performance discrepancies for identical models between EXAONE 3.5 and Kanana technical reports, highlighting reproducibility challenges in Korean NLP evaluations.}
\label{tab:gemma-discrepancy}
\end{table*}

\section{Related Work and Background}

\subsection{Global LLM Benchmarks and Evaluation Frameworks}
The English LLM ecosystem has established mature evaluation frameworks such as Evalverse~\cite{evalverse} and DeepEval~\cite{deepeval}. 
Among them, the \textit{lm-evaluation-harness}~\cite{eval-harness} has been particularly influential, powering the Open LLM Leaderboard and enabling transparent comparison across models.
Its success demonstrates that accessible evaluation tools can democratize model development—allowing small research groups and startups to participate on equal footing.
Similarly, the Chinese framework FlagEval~\cite{FlagEval} has advanced its language ecosystem through an accessibility-first approach that integrates datasets, metrics, and automation into a unified hub.

\subsection{Reproducibility Challenges in Korean LLM Evaluation}
Evaluations of Korean language models show notable inconsistencies across institutions. 
Identical models often report performance gaps of up to ten percentage points on benchmarks such as GSM8K, KMMLU, HumanEval, and MBPP 
when comparing the EXAONE 3.5 report~\cite{exaone} and the Kanana report~\cite{kanana} 
(\autoref{tab:gemma-discrepancy}). 
These discrepancies arise from variations in prompt templates, inference settings, and evaluation criteria, revealing the absence of standardized Korean-specific practices.

Beyond procedural factors, linguistic characteristics of Korean intensify these challenges. 
Its rich morphology, flexible word order, and frequent case-marker omission introduce tokenization and normalization differences that can shift results materially~\cite{jeon2023improvingkoreannlptasks, kim2024doesincompletesyntaxinfluence}. 
Without consistent language-aware handling, results risk reflecting preprocessing artifacts rather than true model competence.

\begin{table*}[ht]
\centering
\small
\renewcommand{\arraystretch}{1.2}
\resizebox{\textwidth}{!}{
\begin{tabular}{l|c|c|c|c|c}
\toprule
\makecell[tl]{\Large \textbf{Feature}}
  & \makecell[tc]{\Large lm-eval-harness\\[2pt] \cite{eval-harness-paper}}
  & \makecell[tc]{\Large OpenAI Evals\\[2pt] \cite{openeval}}
  & \makecell[tc]{\Large Evalverse\\[2pt] \cite{evalverse}}
  & \makecell[tc]{\Large DeepEval\\[2pt] \cite{deepeval}}
  & \makecell[tc]{\Large \textbf{HRET (Ours)}} \\
\midrule
\textbf{Primary Focus} & Academic benchmarks & OpenAI model evaluation & Unified evaluation hub & Unit/regression testing & Korean LLM evaluation \\
\textbf{Developer} & EleutherAI & OpenAI & Upstage & Confident AI & Community Contributors \\
\textbf{Architecture} & Task/Metric modules & Evaluation scripts & Submodule integration & Test case-based & Registry-based plugins \\
\textbf{Built-in Benchmarks} & 60+ general benchmarks & Several public evals & Integrated external benchmarks & Major benchmarks & Korean-focused benchmarks \\
\textbf{Evaluation Metrics} & Traditional metrics & Task-specific scripts & Integrated metrics & 5+ diverse metrics & 5+ diverse metrics \\
\textbf{User Interface} & CLI only & CLI + Dashboard & Slack bot + Web & CLI + Web dashboard & CLI + Python API \\
\textbf{Automation} & Script-based & API-based & No-code automation & CI/CD integration & API-based, CI/CD integration\\
\textbf{Unique Strength} & High reproducibility & OpenAI optimization & Multi-tool integration & Testing-focused & Korean-specific + Registry extensibility \\
\bottomrule
\end{tabular}
}
\caption{Comparison of major evaluation frameworks. HRET uniquely provides registry-based extensibility and Korean-specific benchmarks.}
\label{tab:system-comparison}
\end{table*}

\subsection{Language-Aware Benchmarking for Korean}
While reproducibility issues are procedural in part, they are amplified by the linguistic complexity of Korean, which English-oriented evaluation pipelines often fail to capture~\cite{arnett2024languagemodelsperformworse}. 
As an agglutinative and morphologically rich language, Korean forms words through variable suffixation and inflection. 
Naive subword tokenization (e.g., BPE or WordPiece) frequently misaligns with true morpheme boundaries, reducing evaluation reliability. 
Moreover, flexible word order and frequent case-marker omission make reference matching ambiguous without morphological or syntactic normalization.
Differences in tokenization or spacing can shift scores by several points, conflating model ability with preprocessing artifacts. 
Therefore, a language-aware evaluation framework—featuring morpheme alignment, spacing normalization, and honorific-aware matching—is essential to isolate linguistic competence from implementation variance.

\subsection{Comparison with Existing Systems}
Table~\ref{tab:system-comparison} situates HRET within the broader landscape of evaluation frameworks.
HRET differentiates itself through three design principles:
(1) \textbf{Korean-specific capabilities} such as language-consistency enforcement and morphology-based diagnostics;
(2) \textbf{registry-based extensibility}, enabling rapid integration of new datasets and methods without code restructuring; and 
(3) \textbf{a unified evaluation pipeline} that standardizes fragmented Korean LLM assessment practices.
These choices collectively form the foundation of HRET, described in the following section, which operationalizes these ideas into a reproducible, modular evaluation framework.

\section{\hret Framework}

\textbf{HRET (Haerae Evaluation Toolkit)} addresses the reproducibility and comparability challenges inherent in evaluating Korean large language models (LLMs). 
It provides a standardized yet flexible environment that ensures consistency, transparency, and accessibility while accommodating diverse research and deployment needs. 
Figure~\ref{fig:arch} illustrates its registry-based architecture, which integrates datasets, models, inference strategies, and evaluation methods into a unified pipeline.

HRET targets three primary user groups: (1) researchers seeking reproducible Korean benchmarks; (2) developers requiring reliable evaluation tools for model selection; and (3) practitioners deploying Korean LLMs in production environments.

\begin{table*}[ht]
\centering
\small
\renewcommand{\arraystretch}{1.15}
\setlength{\tabcolsep}{4.8pt}
\begin{tabular}{p{1.8cm} p{1.7cm} p{11.5cm}}
\toprule
\textbf{Category} & \textbf{Benchmark} & \textbf{Description} \\
\midrule
\multirow{3}{*}{\textbf{General}} 
    & \makecell[tl]{HAE-RAE} 
    & Evaluates general Korean language understanding, focusing on vocabulary, reading comprehension, and broad knowledge. \\
    & \makecell[tl]{KMMLU} 
    & Covers 45 knowledge categories (STEM, humanities, applied sciences) in Korean. \\
    & \makecell[tl]{CLIcK} 
    & Tests cultural and linguistic proficiency using 1,995 QA pairs from Korean exams and textbooks. \\
\midrule
\textbf{Dialogue} 
    & \makecell[tl]{KUDGE} 
    & Assesses coherence and contextual appropriateness in Korean dialogue generation. \\
\midrule
\multirow{10}{*}{\textbf{Specialized}} 
    & \makecell[tl]{K2-Eval} 
    & Tests professional and technical Korean terminology. \\
    & \makecell[tl]{HRM8K} 
    & Evaluates mathematical reasoning across 8,011 Korean problems. \\
    & \makecell[tl]{HRMCR} 
    & Multi-step commonsense reasoning using Korean cultural templates. \\
    & \makecell[tl]{HRC} 
    & Korean reasoning challenge with 119 science, mathematics, and quiz tasks. \\
    & \makecell[tl]{KorMedQA} 
    & 7,469 Korean medical licensing MCQs (2012–2024). \\
    & \makecell[tl]{KBL} 
    & Legal reasoning and bar-exam questions in closed-book and RAG settings. \\
    & \makecell[tl]{KoBalt} 
    & Evaluates linguistic competence across syntax, semantics, pragmatics, phonology, and morphology via 700 expert-curated items. \\
    & \makecell[tl]{K-Halu} 
    & Tests hallucination detection across seven knowledge domains from Korean media sources. \\
    & \makecell[tl]{Ko-IFEval} 
    & Korean adaptation of Google’s IFEval dataset for instruction-following assessment. \\
    & \makecell[tl]{Benchhub} 
    & Comprehensive evaluation suite integrating multiple Korean instruction-following and reasoning benchmarks under unified schemas. \\
\bottomrule
\end{tabular}
\caption{Benchmarks currently supported in HRET. Datasets span general, dialogue, and specialized Korean language evaluations.}
\label{tab:benchmarks}
\end{table*}

\subsection{Design Principles}

\textbf{Modular and Scalable Architecture.}
All components—dataset loaders, evaluators, scaling strategies, and backends—are registered modularly, allowing new elements to be added with minimal code modification. 
Parallelized execution using asynchronous I/O and multi-threading enables efficient large-scale benchmarking.

\textbf{Standardized and Reproducible Evaluation.}
HRET unifies data ingestion, inference, and reporting through consistent interfaces and YAML-based configurations, ensuring identical results across hardware setups and institutions.

\textbf{Language-Aware Design.}
The toolkit enforces Korean-only outputs via a language consistency penalty and integrates morphology- and semantics-aware diagnostics to evaluate linguistic robustness beyond surface accuracy.

\subsection{System Architecture}
HRET follows a standardized five-stage workflow—dataset loading, model setup, inference, evaluation, and reporting—covering major Korean benchmarks such as HAE-RAE, KMMLU, KUDGE, HRM8K, and CLIcK (see Section~\ref{sec:benchmark-details}).

\begin{enumerate}
    \item \textbf{Dataset Registry:} Provides consistent dataset formats and a generic loader for custom files.  
    \item \textbf{Backend Registry:} Connects to HuggingFace, vLLM, or OpenAI-compatible APIs through a unified interface that supports both generation and judging modes.  
    \item \textbf{Inference Scaling:} Includes beam search, best-of-$n$ sampling, and majority voting to stabilize outputs and capture diverse reasoning paths.  
    \item \textbf{Evaluation:} Supports four complementary paradigms—%
          (\textit{i}) \textbf{String Matching} for discrete QA and short-form tasks; %
          (\textit{ii}) \textbf{Log-Likelihood Scoring} for multiple-choice and completion-style benchmarks; %
          (\textit{iii}) \textbf{Mathematical Verification} for reasoning and equation-consistency checks; and %
          (\textit{iv}) \textbf{LLM-as-a-Judge} for open-ended or generative responses requiring qualitative assessment. %
          Language-consistency enforcement penalizes non-Korean outputs, and automated answer extraction ensures reliable comparison across reasoning styles.  
    \item \textbf{Reporting:} Aggregates runtime outputs from all registries through a facade-style Evaluator API for unified logging, visualization, and downstream analysis.
\end{enumerate}

\subsection{Supported Benchmarks}
\label{sec:benchmark-details}

HRET provides unified evaluation support for major Korean language benchmarks spanning diverse domains and task types. 
Its registry-based architecture enables seamless integration of these benchmarks with consistent preprocessing, evaluation protocols, and result formatting.

The collection covers three major categories: 
\textbf{(1) General} benchmarks such as HAE-RAE~\cite{haerae}, KMMLU~\cite{kmmlu}, and CLIcK~\cite{click}, which assess broad Korean language understanding and cultural knowledge; 
\textbf{(2) Dialogue} benchmarks like KUDGE~\cite{kudge}, which evaluate conversational coherence and contextual appropriateness; and 
\textbf{(3) Specialized} benchmarks that target domain-specific reasoning and professional knowledge, including K2-Eval~\cite{kudge}, HRM8K~\cite{ust}, HRMCR~\cite{hrmcr}, 
KorMedQA~\cite{kormedqa}, KBL~\cite{kbl}, KoBalt~\cite{kobalt}, K-Halu~\cite{khalu}, Ko-IFEval~\cite{lee2024ifevalko}, and Benchhub~\cite{benchhub}. 
This comprehensive coverage enables holistic evaluation across linguistic, reasoning, and professional knowledge domains.

Table~\ref{tab:benchmarks} provides detailed descriptions of the supported benchmarks organized by evaluation focus.

\subsection{Advanced and Extensible Pipelines}
Beyond standard evaluation, HRET’s \texttt{PipelineRunner} enables advanced workflows including multi-model comparisons, ensemble scoring, and cascading evaluations where one model’s outputs feed another. 
A high-level decorator interface inspired by DeepEval simplifies experiment definition while maintaining compatibility with the underlying pipeline:

\begin{lstlisting}
import llm_eval.hret as hret

@hret.evaluate(dataset="kmmlu", model="huggingface",
               evaluation_method="string_match")
def my_model(x: str) -> str:
    return model.generate(x)
result = my_model()
print(result.metrics["accuracy"])
\end{lstlisting}

Both interfaces are interoperable and integrate seamlessly with MLOps tools (e.g., MLflow, W\&B) for experiment tracking and result persistence.

\subsection{Korean-Specific Diagnostic Analyses}
HRET extends beyond conventional accuracy metrics by incorporating morphology-aware analyzers and semantic diagnostics to capture language-specific phenomena. 
It provides four diagnostic views: (1) \textbf{Performance Overview} with detailed accuracy and prediction distributions; (2) \textbf{Subset Analysis} across domains (e.g., reasoning, culture, comprehension); (3) \textbf{Lexical Diversity} via Type–Token Ratio (TTR) between correct and incorrect responses; and (4) \textbf{Error Pattern Detection} through systematic keyword omission tracking. 
These analyses expose morphological and semantic weaknesses invisible to surface metrics, guiding targeted model refinement.

HRET also defines three extended task families—(i) instruction following (schema adherence and refusal compliance), (ii) long-context reasoning (cross-chunk consistency), and (iii) hallucination stress tests (factuality under distractors). 
Each family includes metric adapters and Korean honorific-aware judge prompts to ensure culturally and linguistically appropriate assessment.

\subsection{Usage and Reproducibility}
HRET supports both command-line and Python interfaces for reproducible benchmarking. 
A minimal evaluation example is shown below, with extended configurations and templates available in the project repository.

\begin{lstlisting}
python -m llm_eval.evaluator \
  --model huggingface \
  --dataset haerae_bench \
  --evaluation_method string_match
\end{lstlisting}

\section{System Evaluation}

\subsection{Reproducibility Validation}
\label{sec:reproducibility}
To evaluate HRET's effectiveness in addressing reproducibility challenges, we conducted cross-institutional experiments across multiple hardware configurations and backend implementations. Three representative models—Qwen2.5-3B~\cite{qwen25}, Qwen3-8B~\cite{qwen3}, and Llama-3.2-3B-Instruct~\cite{llama_3.2_3b_instruct}—were tested on Korean benchmarks using log-likelihood evaluation across four heterogeneous environments differing in CPU architecture, GPU type, and memory capacity.

\begin{table}[ht]
\centering
\small
\begin{tabular}{lcc}
\toprule
\textbf{Model} & \textbf{Max Variation} & \textbf{Hardware Configs} \\
\midrule
Qwen2.5-3B      & 0.0016 (0.56\%) & 4 setups \\
Qwen3-8B        & 0.0028 (1.06\%) & 4 setups \\
Llama-3.2-3B    & 0.0015 (0.62\%) & 4 setups \\
\bottomrule
\end{tabular}
\caption{Hardware robustness summary across environments.}
\end{table}

Table~\ref{tab:hardware-details} provides the full per-configuration results. Across server-grade (A100, H100) and workstation-grade (RTX A6000, RTX 3090) environments, scores remain nearly identical, confirming HRET's hardware-agnostic consistency.

\begin{table*}[ht]
\centering
\tiny
\renewcommand{\arraystretch}{0.9}
\resizebox{0.9\textwidth}{!}{
\begin{tabular}{llccc}
\toprule
\textbf{Model} & \textbf{Hardware Setup} & \textbf{HAE-RAE} & \textbf{CLIcK} & \textbf{Batch Size} \\
\midrule
\multirow{4}{*}{\shortstack[l]{Qwen2.5-3B}}
& AMD EPYC 74F3 + A100×2        & 0.2847 & 0.3388 & 64 \\
& AMD EPYC 9354 + RTX A6000     & 0.2847 & 0.3388 & 2  \\
& Intel i9-13900K + RTX A6000   & 0.2847 & 0.3388 & 2  \\
& Intel Xeon + A100-40GB        & 0.2863 & 0.3378 & 64 \\
\midrule
\multirow{4}{*}{\shortstack[l]{Qwen3-8B}}
& AMD EPYC 74F3 + A100×2        & 0.2645 & 0.3083 & 64 \\
& AMD EPYC 9354 + RTX A6000     & 0.2645 & 0.3083 & 1--2 \\
& Intel i9-13900K + RTX A6000   & 0.2645 & 0.3083 & 2 \\
& Intel Xeon + A100-40GB        & 0.2673 & 0.3083 & 64 \\
\midrule
\multirow{4}{*}{\shortstack[l]{Llama-3.2-3B}}
& AMD EPYC 74F3 + A100×2        & 0.2436 & 0.2922 & 64 \\
& AMD EPYC 7452 + RTX 3090      & 0.2437 & 0.2922 & 2 \\
& Intel i9-13900K + RTX A6000   & 0.2445 & 0.2922 & 2 \\
& Intel Xeon + A100-40GB        & 0.2447 & 0.2907 & 64 \\
\bottomrule
\end{tabular}}
\caption{Reproducibility validation across hardware configurations using HuggingFace backend.}
\label{tab:hardware-details}
\end{table*}

These results confirm that HRET effectively eliminates the 1--10 percentage point discrepancies previously observed between institutions (cf.\ Table~\ref{tab:gemma-discrepancy}). Within each backend, results remain statistically identical ($\sigma<0.003$), confirming that HRET ensures reproducible evaluation even under varied computational environments.

\begin{table}[ht]
\centering
\small
\renewcommand{\arraystretch}{1.1}
\begin{tabular}{lcc}
\toprule
\textbf{Hardware Configuration} & \textbf{HAE-RAE} & \textbf{CLIcK} \\
\midrule
AMD EPYC 74F3 + H100          & 0.1806 & 0.1343 \\
Intel Xeon + A100-80GB        & 0.1788 & 0.1298 \\
Intel Xeon + H200             & 0.1802 & 0.1308 \\
\midrule
\textbf{Standard Deviation}   & \textbf{0.0009} & \textbf{0.0023} \\
\bottomrule
\end{tabular}
\caption{Backend robustness using vLLM for Qwen2.5-3B.}
\label{tab:backend-comparison}
\end{table}

\subsection{Config-Driven Reproducibility via YAML}
To facilitate exact replication across institutions, HRET employs standardized, shareable YAML configurations. For all integrated benchmarks, published scores can be reproduced with less than 1\% deviation by matching only configuration keys—such as prompt templates, decoding parameters, and judge settings—while keeping dataset splits fixed:

\begin{lstlisting}
# hret_config.yaml
default_dataset: "kmmlu"
default_model: "huggingface"
default_split: "test"
default_evaluation_method: "string_match"
batch_size: 32
max_workers: 4
\end{lstlisting}

\begin{lstlisting}
import llm_eval.hret as hret
hret.load_config("hret_config.yaml")
res = hret.quick_eval(my_model_function, dataset="kmmlu")
\end{lstlisting}

This configuration discipline eliminates hidden variability, enabling reproducible “paper-accurate” replication across teams and hardware.

\subsection{Multi-Method Evaluation Analysis}
Korean tasks vary widely in linguistic and reasoning complexity, requiring different evaluation strategies. HRET's multi-method framework enables appropriate metric selection per task, balancing accuracy and interpretability.

Table~\ref{tab:multi-method} compares evaluation methods for EXAONE-3.5-2.4B~\cite{exaone} across two benchmarks. Results reveal method sensitivity in reasoning tasks and stability in linguistic ones.

\begin{table}[h]
\centering
\small
\setlength{\tabcolsep}{3pt}
\renewcommand{\arraystretch}{0.96}
\resizebox{1\linewidth}{!}{
\begin{tabular}{lccc}
\toprule
\textbf{Benchmark} & \textbf{String-Match} & \textbf{Math-Verify} & \textbf{LLM-as-judge} \\
\midrule
HRM8K & 2.21  & 14.11 & 17.60 \\
CLIcK & 27.31 & 28.67 & 27.76 \\
\bottomrule
\end{tabular}
}
\caption{Comparison of evaluation methods for EXAONE-3.5-2.4B using GPT-4o (2024-11-20) as the LLM-as-judge.}
\label{tab:multi-method}
\end{table}

Mathematical reasoning (HRM8K) benefits substantially from semantic evaluation—scores increase from 2.2\% (string-match) to 17.6\% (LLM-as-judge)—whereas linguistic and cultural QA (CLIcK) remain robust across metrics (27–29\%). These findings validate HRET's ability to align evaluation method with task type.

\subsection{Test-Time Scaling Techniques}
To further enhance evaluation robustness, HRET integrates test-time scaling strategies that average or aggregate predictions from multiple inference runs. The framework supports three methods: \textbf{Best-of-N Sampling}, \textbf{Self-Consistency} via majority voting, and \textbf{Beam Search} for structured decoding. These can be combined with any evaluation metric, including LLM-as-a-judge assessments.

\begin{table}[h]
\centering
\small
\setlength{\tabcolsep}{2pt}
\renewcommand{\arraystretch}{0.95}
\begin{tabular}{lcccc}
\toprule
\textbf{Model} & \textbf{Benchmark} & \textbf{n=1} & \textbf{n=3} & \textbf{n=5} \\
\midrule
EXAONE-3.5-2.4B         & \multirow{2}{*}{HRC} & 3.37  & 5.04  & 9.24 \\
Qwen2.5-3B-Instruct     &                       & 11.76 & 12.60 & 13.44 \\
\bottomrule
\end{tabular}
\caption{Test-time scaling with LLM-as-a-judge (GPT-4o-2024-11-20) on HRC.}
\label{tab:test-time-scaling}
\end{table}

Increased sampling significantly improves performance—EXAONE-3.5-2.4B gains 2.7$\times$ at $n=5$ compared to single-shot inference—confirming the benefit of stochastic aggregation for reasoning-intensive Korean tasks.

\begin{table}[ht]
\centering
\small
\setlength{\tabcolsep}{2.5pt}
\renewcommand{\arraystretch}{1.0}
\begin{tabular}{lcc}
\toprule
\textbf{Metric} & \textbf{HRET} & \textbf{lm-eval-harness} \\
\midrule
\multicolumn{3}{c}{\textbf{HAE-RAE Bench}} \\
\midrule
Execution Time (sec) & 114.93 & 142.44 \\
Max RAM (MB) & 80.96 (<1\%) & 54.26 (<1\%) \\
Max CPU (\%) & 525.84 & 100.0 \\
\midrule
\multicolumn{3}{c}{\textbf{KMMLU}} \\
\midrule
Execution Time (sec) & 303.93 & 221.04 \\
Max RAM (MB) & 104.81 (<1\%) & 78.76 (<1\%) \\
Max CPU (\%) & 458.90 & 199.0 \\
\bottomrule
\end{tabular}
\caption{Performance comparison on AMD EPYC 74F3 64-Core + NVIDIA H100 with 193GB RAM. Values averaged across 3 runs.}
\label{tab:performance-comparison}
\end{table}

\subsection{Performance Comparison with Existing Tools}
\label{appendix:performance-comparison}
Table~\ref{tab:performance-comparison} compares HRET's computational performance with lm-evaluation-harness using Qwen2.5-3B on identical hardware configurations.

HRET shows mixed performance: faster on HAE-RAE Bench but slower on KMMLU. Higher RAM usage stems from loading datasets as editable objects to enable real-time filtering and experimental modifications through our system. Higher CPU utilization (up to 525\%) reflects multi-threading optimization. Both frameworks use <1\% of total system memory, indicating efficient resource utilization despite HRET's flexibility-oriented design.

Across all experiments, HRET demonstrates (i) near-zero variance across hardware and backends, (ii) configuration-based reproducibility within 1\%, and (iii) flexible adaptation to diverse evaluation paradigms. 
Together, these results validate HRET as a robust, extensible foundation for reproducible and language-aware Korean LLM benchmarking.

\subsection{Extensibility}
HRET’s registry-based design ensures seamless extensibility. New components—%
\textbf{Models}, \textbf{Datasets}, \textbf{Evaluators}, and \textbf{Scaling Methods}—%
can be added through a single decorator (e.g., \verb|@register_model|), automatically integrating with the existing runtime without additional configuration.

This plug-and-play approach encourages community contributions and supports continuous evolution of the framework alongside emerging Korean benchmarks and inference paradigms, while maintaining full backward compatibility.

\section{Real-World Adoption}
Beyond academic evaluation, HRET has been adopted in production environments across the Korean NLP ecosystem. The Horangi Korean LLM Leaderboard\footnote{\url{https://wandb.ai/wandb-korea/korean-llm-leaderboard/reports/Horangi-LLM---Vmlldzo3MzIyNDE2}}, a community-driven leaderboard hosted on Weights \& Biases, uses HRET as its evaluation backbone to ensure standardized and reproducible model comparison. Additionally, the Haerae Reasoning Challenge (HRC) officially employs HRET to score all participant submissions under standardized protocols, ensuring fairness and transparency in competitive evaluations. Authors' affiliated organizations, including OneLineAI, have also integrated HRET into their model validation and post-training evaluation workflows, demonstrating the framework's practical utility in deployment scenarios.
This real-world adoption validates HRET's reliability beyond controlled research settings, confirming its value as evaluation infrastructure for both the research community and industry practitioners.

\section{Conclusion}
We introduce HRET, a unified evaluation framework that addresses critical reproducibility challenges in Korean LLM evaluation. Through its registry-based architecture and standardized protocols, HRET significantly improves evaluation consistency while supporting diverse assessment strategies.
HRET integrates major Korean benchmarks with multiple inference backends and evaluation methods, enabling researchers to select task-appropriate strategies while maintaining reproducibility. The framework's extensible design and community-driven development ensure it evolves with the Korean LLM landscape, providing a sustainable foundation for reproducible assessment.

\section*{Limitations}
While HRET provides a standardized and reproducible evaluation framework for Korean LLMs, several limitations remain. 
First, although it integrates most major Korean benchmarks, coverage in specialized domains~\cite{twice, kopiqa} is still limited, and the current framework does not yet support vision-language model (VLM) benchmarks~\cite{choi2026users}, which are increasingly important for multimodal Korean evaluation. Second, the language consistency enforcement may struggle with mixed-language or code-switched outputs common in technical and multilingual contexts. 
Third, test-time scaling techniques require considerable computational resources, which may limit accessibility for smaller institutions or resource-constrained environments. 
In addition, current metrics—particularly string matching—do not fully capture Korean's morphological richness and honorific nuances, and the framework's accuracy ultimately depends on the quality of integrated benchmarks. 
We are actively addressing these challenges through ongoing development and invite community contributions to further enhance HRET's scope and robustness.

\section*{Ethics Statement}
HRET aims to improve reproducibility and fairness in Korean LLM evaluation. However, we acknowledge several ethical considerations: (1) standardized evaluation frameworks may inadvertently favor certain model architectures or training approaches; (2) benchmark selection reflects our research perspectives and may not capture all aspects of Korean language competency; (3) language consistency enforcement, while promoting Korean language assessment, may disadvantage multilingual models in certain contexts. We encourage the community to contribute diverse benchmarks and evaluation methods to ensure broad representation of Korean language capabilities. All integrated datasets respect original licensing and guidelines.

% ---------------- References ----------------
\bibliographystyle{lrec2026-natbib}
\bibliography{custom}

\end{document}